\documentclass[aps,prl,amsmath,amssymb,twocolumn,showpacs,superscriptaddress,final,floatfix]{revtex4-1}
\usepackage{graphicx}	
\usepackage{color}	
\usepackage[normalem]{ulem}
\usepackage{hyperref}
\hypersetup{
  colorlinks,
  citecolor=blue,
  linkcolor=blue,
  urlcolor=blue}

\newcommand{\BFA}{BaFe$_2$As$_2$}

\newcommand{\ie}{{\em i.e.}}

\begin{document}

\title{Anisotropic quasiparticle coherence in nematic BaFe$_2$As$_2$ studied with strain-dependent ARPES}

\author{H. Pfau}
\email{hpfau@stanford.edu}
\affiliation{Advanced Light Source, Lawrence Berkeley National Laboratory, Berkeley, California 94720, USA}
\affiliation{Geballe Laboratory for Advanced Materials, Department of Applied Physics, Stanford University, Stanford, California 94305, USA}
\affiliation{Stanford Institute of Materials and Energy Science, SLAC National Accelerator Laboratory, Menlo Park, California 94025, USA}
\author{S. D. Chen}
\affiliation{Geballe Laboratory for Advanced Materials, Department of Applied Physics, Stanford University, Stanford, California 94305, USA}
\author{M. Hashimoto}
\affiliation{Stanford Synchrotron Radiation Lightsource, SLAC National Accelerator Laboratory, Menlo Park, California 94025, USA
}
\author{N. Gauthier}
\affiliation{Geballe Laboratory for Advanced Materials, Department of Applied Physics, Stanford University, Stanford, California 94305, USA}
\affiliation{Stanford Institute of Materials and Energy Science, SLAC National Accelerator Laboratory, Menlo Park, California 94025, USA}
\author{C. R. Rotundu}
\affiliation{Stanford Institute of Materials and Energy Science, SLAC National Accelerator Laboratory, Menlo Park, California 94025, USA}
\author{J. C. Palmstrom}
\affiliation{Geballe Laboratory for Advanced Materials, Department of Applied Physics, Stanford University, Stanford, California 94305, USA}
\affiliation{Pulsed Field Facility, National High Magnetic Field Laboratory, Los Alamos National Laboratory, Los Alamos, New Mexico 87545, USA}
\author{I. R. Fisher}
\affiliation{Stanford Institute of Materials and Energy Science, SLAC National Accelerator Laboratory, Menlo Park, California 94025, USA}
\affiliation{Geballe Laboratory for Advanced Materials, Department of Applied Physics, Stanford University, Stanford, California 94305, USA}
\author{S.-K. Mo}
\affiliation{Advanced Light Source, Lawrence Berkeley National Laboratory, Berkeley, California 94720, USA}
\author{Z.-X. Shen}
\email{zxshen@stanford.edu}
\affiliation{Stanford Institute of Materials and Energy Science, SLAC National Accelerator Laboratory, Menlo Park, California 94025, USA}
\affiliation{Department of Physics, Stanford University, Stanford, California 94305 , USA}
\affiliation{Geballe Laboratory for Advanced Materials, Department of Applied Physics, Stanford University, Stanford, California 94305, USA}
\author{D. Lu}
\email{dhlu@slac.stanford.edu}
\affiliation{Stanford Synchrotron Radiation Lightsource, SLAC National Accelerator Laboratory, Menlo Park, California 94025, USA
}

\date{\today}


\begin{abstract}

The hallmark of nematic order in iron-based superconductors is a resistivity anisotropy but it is unclear to which extent quasiparticle dispersions, lifetimes and coherence contribute. While the lifted degeneracy of the Fe $d_{xz}$ and $d_{yz}$ dispersions has been studied extensively, only little is known about the two other factors. Here, we combine {\it{in situ}} strain tuning with ARPES and study the nematic response of the spectral weight in \BFA. The symmetry analysis of the ARPES spectra demonstrates that the $d_{xz}$ band gains quasiparticle spectral weight compared to the $d_{yz}$ band for negative antisymmetric strain $\Delta \epsilon_{yy}$ suggesting the same response inside the nematic phase. Our results are compatible with a different coherence of the $d_{xz}$ and $d_{yz}$ orbital within a Hund's metal picture. We also discuss the influence of orbital mixing.
\end{abstract}

\maketitle


\section{Introduction}

A central topic in the study of iron-based superconductors (FeSC) is the interplay of superconductivity and nematic order. It is therefore important to clarify the origin and properties of nematic order. Nematicity is an electronic instability that breaks the rotational symmetry and induces an anisotropy for the spin, orbital and lattice degrees of freedom. \cite{fernandes_2014,paglione_2010,johnston_2010}. In general, the nematic $C_4$ symmetry breaking can occur in different $B_{2g}$ channels, which either involve on-site interactions called ferro-orbital order or interactions between neighboring Fe-atoms called bond order. Several experimental results, in particular the sign change of the nematic band splitting across the Brillouin zone \cite{suzuki_2015,pfau_2019_prl} sketched in Fig.~\ref{Fig:intro}(b), cannot be accounted for by ferro-orbital order. Therefore, interatomic interactions are considered responsible for the nematic order. Even though nematicity has been studied extensively, there is little consensus regarding the underlying microscopic origin of these interactions and both spin and orbital degrees of freedom have been suggested \cite{fernandes_2014,chubukov_2016,beak_2014,boehmer_2014}.

The effect of nematicty on orbital degrees of freedom has mainly been characterized by the energy shifts of the $d_{xz}$ and $d_{yz}$ bands studied with angle-resolved photemission spectroscopy (ARPES) \cite{yi_2011_pnas,jensen_2011,kim_2011,yi_2012_njp,nakayama_2014,watson_2015,suzuki_2015,fedorov_2016,zhang_2016,watson_2017_njp,pfau_2019_prb,fedorov_2019_prb,pfau_2019_prl,yi_2019_prx}.
However, anisotropies in the quasiparticle line width and more generally in the electronic spectral weight distribution are expected as well if interatomic interactions drive nematicity. For example, the coupling of electronic quasiparticles to anisotropic magnetic fluctuations influences the inelastic scattering rate and will lead to an anisotropic line width \cite{fernandes_2011,fanfarillo_2016}. Furthermore, FeSC are considered Hund's metals \cite{haule_2009,johannes_2009,medici_2014}. Lifting the degeneracy of the $d_{xz}$ and $d_{yz}$ orbitals within this framework is expected to change their quasiparticle coherence, which is measured by the amount of coherent spectral weight. Nematicity will thus lead to an anisotropic coherent spectral weight \cite{fanfarillo_2017,yu_2018}. Studying the changes in quasiparticle lifetime and coherence will ultimately also evolve our understanding of the resistivity anisotropy, which is one of the hallmarks of the nematic order.

Recent ARPES measurements provided first indications for a temperature-dependent change of spectral weight within the nematic phase \cite{cai_2020}. Additionally, optical spectroscopy provided evidence for anisotropic Drude weights and scattering rates upon entering the nematic state \cite{homes_2020_prb,chinotti_2018,mirri_2015,schuett_2016}. However, the momentum and orbital integrated nature of optical spectroscopy prevents a deeper, microscopic understanding necessary to draw conclusion about the microscopic mechanism behind nematicity.

Therefore, we study the influence of nematicity on the quasiparticle spectral weight using ARPES. We use {\em in-situ} tunable uniaxial stress to extract the nematic $B_{2g}$ contribution to the spectral weight change in response to antisymmetric strain. We choose the prototype FeSC \BFA~and measure its strain response at a temperature above the nematic and magnetic phase transitions, which are located in close proximity. This strategy prevents magnetism from obstructing the nematic signatures. Our data demonstrate that the $d_{xz}$ ($d_{yz}$) orbital gains (looses) quasiparticle spectral weight due to nematic order as sketched in Fig.~\ref{Fig:intro}(c). We discuss our results within a picture of Hund's metal physics and illustrate the influence of orbital mixing.


\section{Methods}

\begin{figure}
\includegraphics[width=\columnwidth]{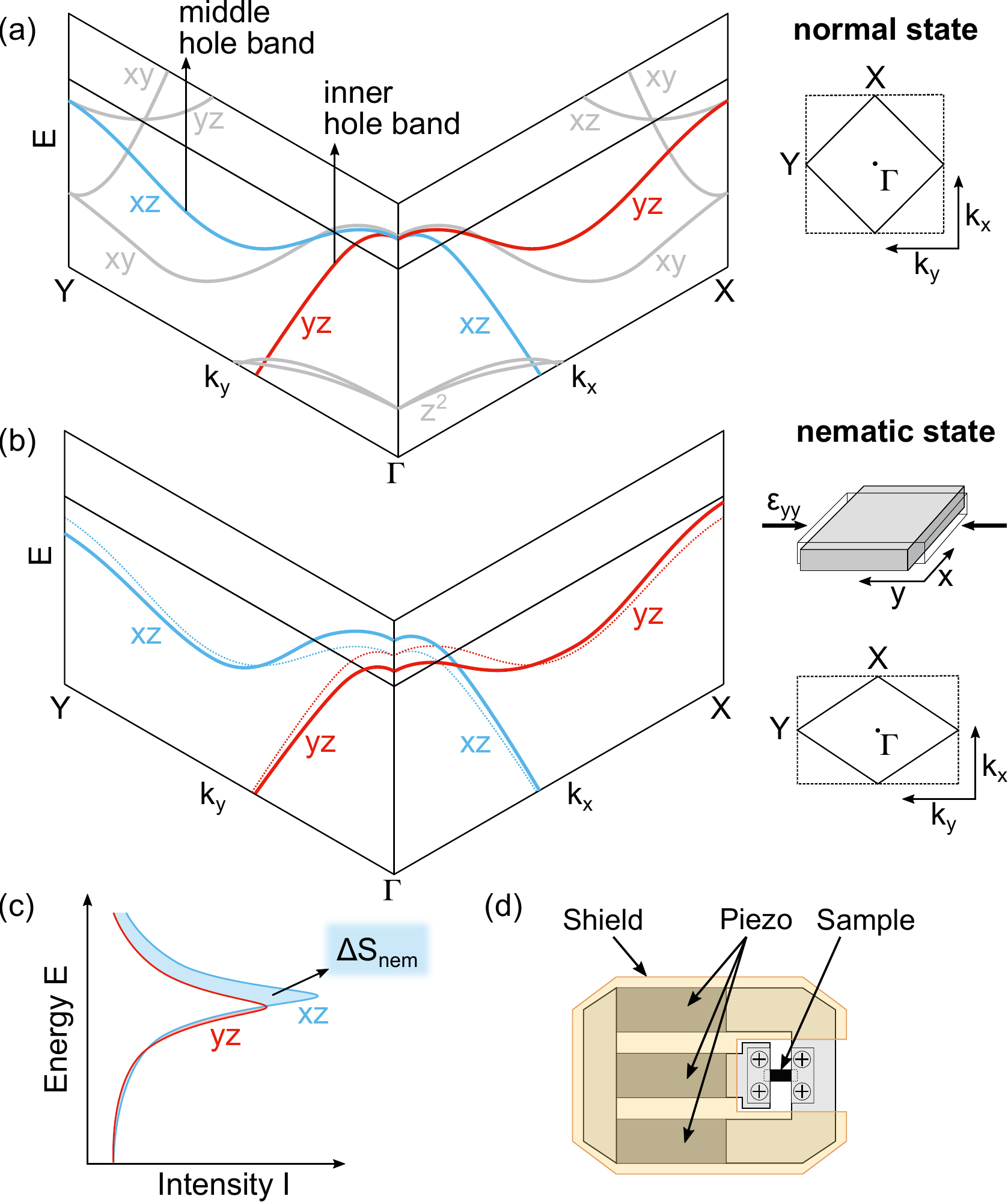}
\caption{
Changes in the spectral function due to nematicity. (a) Sketch of bandstructure and BZ (solid: 2Fe BZ, dashed 1Fe BZ) in the normal state. (b) Bandstructure and BZ in the nematic state. Sketch of the sample indicates the corresponding deformation due to uniaxial strain $\epsilon_{yy}$ or nematicty, respectively. Dashed lines in bandstructure indicate dispersion in normal state for comparison. (c) Anisotropy of the coherent spectral weight $\Delta S_\mathrm{nem}$ between $d_{xz}$ and $d_{yz}$ orbitals inside the nematic state. (d) Sketch of the uniaxial strain device.
}
\label{Fig:intro}
\end{figure}

High quality single crystals of \BFA~were grown using a self-flux method \cite{chu_2009,wang_2009,rotundu_2010}. We studied \BFA~at $160\,\mathrm{K}$, \ie~above the nematic and magnetic phase transition temperatures of approximately 140\,K. A strain device with three piezoelectric stacks as sketched in Fig.~\ref{Fig:intro}(d) and described in Ref.~\cite{pfau_2019_prl} is used to apply an {\it{in situ}} tuneable uniaxial pressure along the in-plane Fe-Fe bond directions, which we call $y$ without loss of generality (Fig.~\ref{Fig:intro}(b)). The resulting coordinate system is $45^\circ$ rotated with respect to the tetragonal a and b axes. The uniaxial pressure will result in a symmetric and antisymmetric strain response. The latter has the same $B_{2g}$ symmetry as the nematic order. A symmetry analysis under uniaxial pressure therefore allows us to determine spectral weight changes due to nematicity. Using a nonthermal tuning parameter also avoids contributions due to temperature-induced changes in coherence observed in many FeSC \cite{yi_2013_prl,yi_2015_NatCom,pu_2016_prb}. 

We compare spectra taken with compressive and tensile pressure that correspond to +(-)90\,V applied to the center(outer) piezoelectric stacks and {\it{vice versa}}. We applied the same uniaxial pressure for the two orthogonal momentum directions we studied. A strain gauge was used to estimate the strain between both settings to be $\Delta l/l \approx 0.16$\%.  ARPES measurements were performed at SSRL beamline 5-2 with an energy and angular resolution of 12\,meV and 0.1$^\circ$. The samples are cleaved {\it{in situ}} with a base pressure below $5\cdot 10^{-11}$\,Torr. We use a photon energy of 47\,eV, which probes a $k_z$ close to the $\Gamma$ point of the Brillouin zone (BZ) \cite{brouet_2009}. We confirmed,
that a metallic shielding prevents the high voltage of the piezoelectric stacks to alter the ARPES measurement. Standard detector anisotropies are characterized and removed using separate measurements on polycrystalline gold. The presented ARPES spectra are normalized by the photon beam current and divided by a Fermi-Dirac distribution convoluted with the instrument resolution.  

\begin{figure}
\includegraphics[width=\columnwidth]{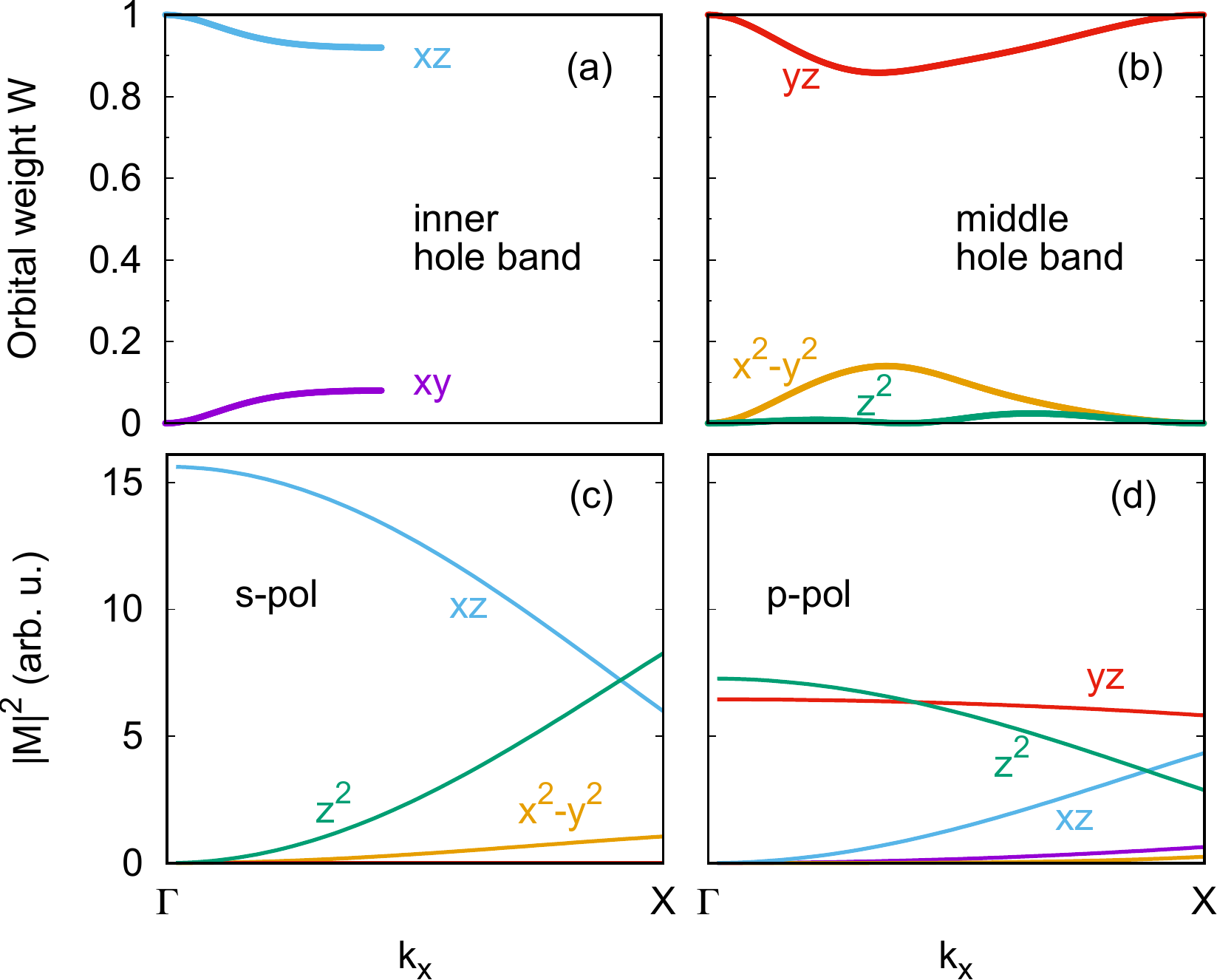}
\caption{
(a,b) Tight-binding calculations of the orbital content of the inner and middle hole bands along $k_x$. (c,d) Matrix elements of the Fe $3d$ orbitals calculated along $k_x$ for the experimental geometry and a photon energy of 47\,eV. The role of $d_{xz}$ and $d_{yz}$ flip along $k_y$.
}
\label{Fig:matrix}
\end{figure}

We focus our study on the inner and middle hole band shown in Fig.~\ref{Fig:intro}(a,b). The calculation of a 10-band tight-binding model with parameters from Ref.~\onlinecite{eschrig_2009_prb} reveals a predominantly $d_{xz}$ and $d_{yz}$ orbital character of these two bands as depicted in Fig.~\ref{Fig:matrix}(a,b). To selectively probe them, we choose linear horizontal (p-pol) and linear vertical (s-pol) polarized light. The corresponding photoemission dipole matrix elements $M = \langle f | \vec{A}\cdot\vec{r} | i \rangle$ for the experimental parameters are shown in Fig.~\ref{Fig:matrix}(c,d). They were calculated in the length gauge using the approximation of a free electron final state and Fe$3d$ hydrogen-like wave functions as initial states \cite{goldberg_1978,gadzuk_1975}.

\section{Experimental Results}

\begin{figure*}
\includegraphics[width=\textwidth]{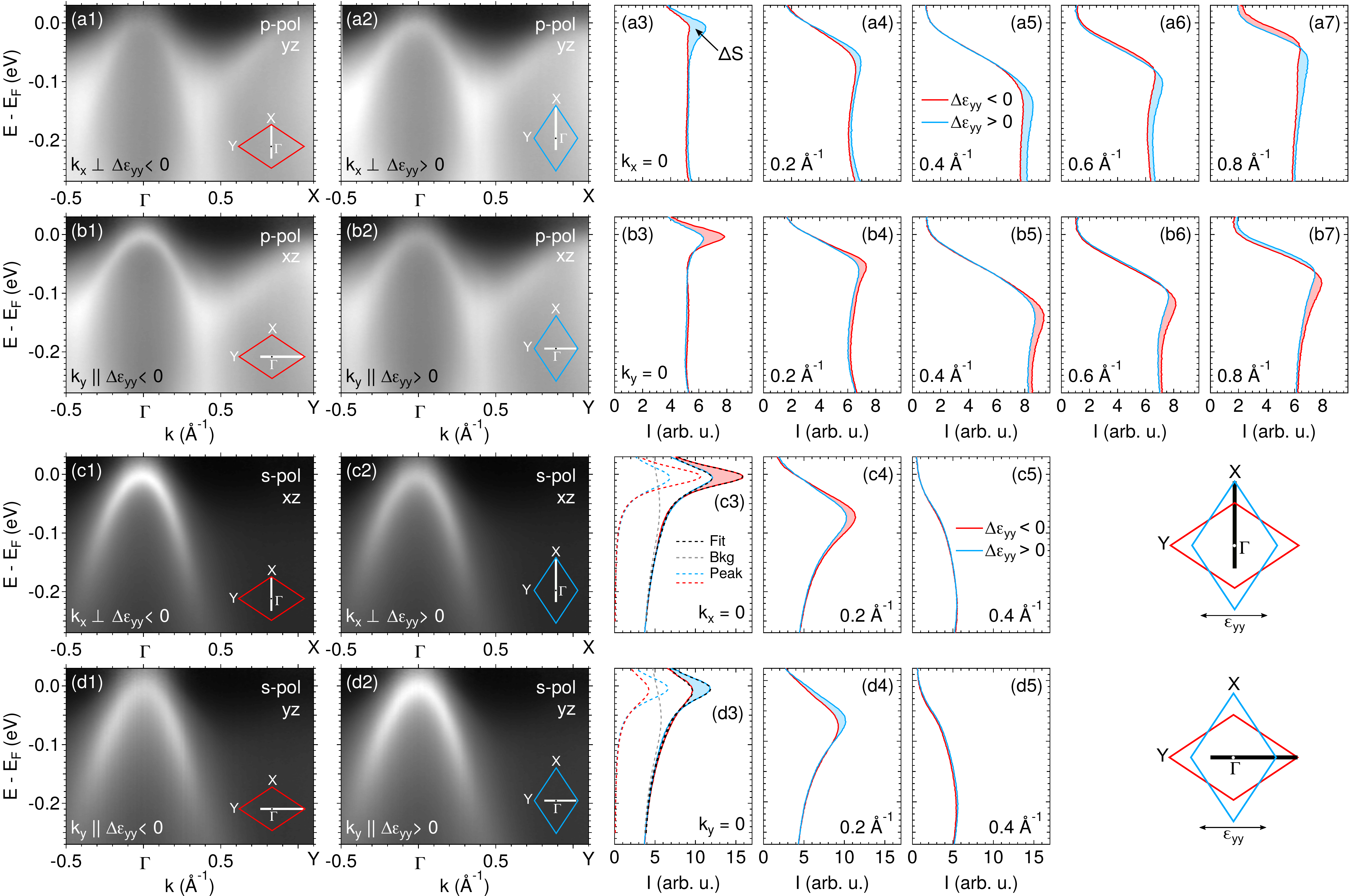}
\caption{
ARPES of strained \BFA. (a) Spectra taken along $k_x$ under compression ($\epsilon_{yy}<0$) and tension ($\epsilon_{yy}>0$) in p-pol together with selected EDCs (a3)-(a7). This configuration probes $d_{yz}$ character as indicated in (a1,a2). Blue (red) shading between the EDCs indicates a larger (smaller) intensity under tension $\Delta\epsilon_{yy}>0$. The shaded area defines the change in spectral weight $\Delta S$ according to Eqn.~\ref{eqn:dS}. (b) Same as (a) for the orthogonal momentum direction $k_y$ probing $d_{xz}$ character. (c,d) Same as (a,b) for s-pol light. Dashed lines in (c3) and (d3) represent fits (black) with their peak (red, blue) and background (gray) contributions. The insets and the sketches on the right bottom illustrate the strain and momentum directions with respect to the BZ.
}
\label{Fig:Ba122}
\end{figure*}

Figure \ref{Fig:Ba122} shows the ARPES spectra on strained \BFA. We compare the spectra along $k_x$ and $k_y$ for p-pol light in Fig.~\ref{Fig:Ba122}(a,b) and for s-pol light in Fig.~\ref{Fig:Ba122}(c,d). The calculated photoemission matrix elements demonstrate the widely exploited selection rules: Close enough to $\Gamma$, s-pol (p-pol) light photoemits electrons with $d_{xz}$ ($d_{yz}$) character along $k_x$ and electrons with $d_{yz}$ ($d_{xz}$) character along $k_y$. Therefore, we see the inner hole band in s-pol and the middle hole band in p-pol. When $d_{xz}$ and $d_{yz}$ mix due to spin-orbit coupling around $\Gamma$, we will still be sensitive to either one of the orbitals in a certain polarization and can therefore distinguish their overall spectral weight contribution. 

Away from $\Gamma$, p-pol light photoemits both $d_{xz}$ and $d_{yz}$ electrons and we pick up spectral weight from the electron bands at the BZ corner. To avoid mixing signatures of hole and electron bands and to maintain clear selection rules, we will restrict the analysis throughout this paper to $k_{x,y}<0.8\,\mathrm{\AA{}}$. The inner and middle hole bands also contain a sizable $d_{xy}$ and $d_{x^2-y^2}$ character, respectively. However, the matrix elements for these orbitals are essentially zero. Therefore, our measurements will be sensitive exclusively to the $d_{xz,yz}$ contribution to the middle and inner hole band.

The spectra in Fig.~\ref{Fig:Ba122}(a1,a2) and the corresponding energy distribution curves (EDCs) in Fig.~\ref{Fig:Ba122}(a3-a7) show the pressure-induced spectral changes. On one hand, we observe a shift of the band position. We use these spectra in Ref.~\cite{pfau_2019_prl} to analyze the band shifts in detail. On the other hand, the spectral weight under compression ($\epsilon_{yy}<0$) is smaller than under tension ($\epsilon_{yy}>0$). The spectral weight change is independent of the sign of the binding energy shift and it stays nonzero even for zero shift. This observation is different from the Fermi liquid expectation that the quasiparticle spectral weight is expected to increases closer to the Fermi level. Rotating the momentum direction changes the sign in the spectral weight response shown in Fig.~\ref{Fig:Ba122}(b). Hence we observe a strain-induced antisymmetric spectral weight change due to nematicity. The spectra taken with s-pol light in Fig.~\ref{Fig:Ba122}(c,d) show the same effect on the inner hole band but with opposite sign compared to p-pol. The orbital character swaps between inner and middle hole band. Therefore, the spectral weight response is tied to the orbital character and not to a specific band. A negative antisymmetric strain $\Delta\epsilon_{yy} < 0$ (red traces in Fig.~\ref{Fig:Ba122}) corresponds to the lattice distortion $a>b$ inside the nematic phase. We therefore conclude that \BFA~gains $d_{xz}$ and looses $d_{yz}$ spectral weight due to nematicity as sketched in Fig.~\ref{Fig:intro}(c). 

\begin{figure}
\includegraphics[width=\columnwidth]{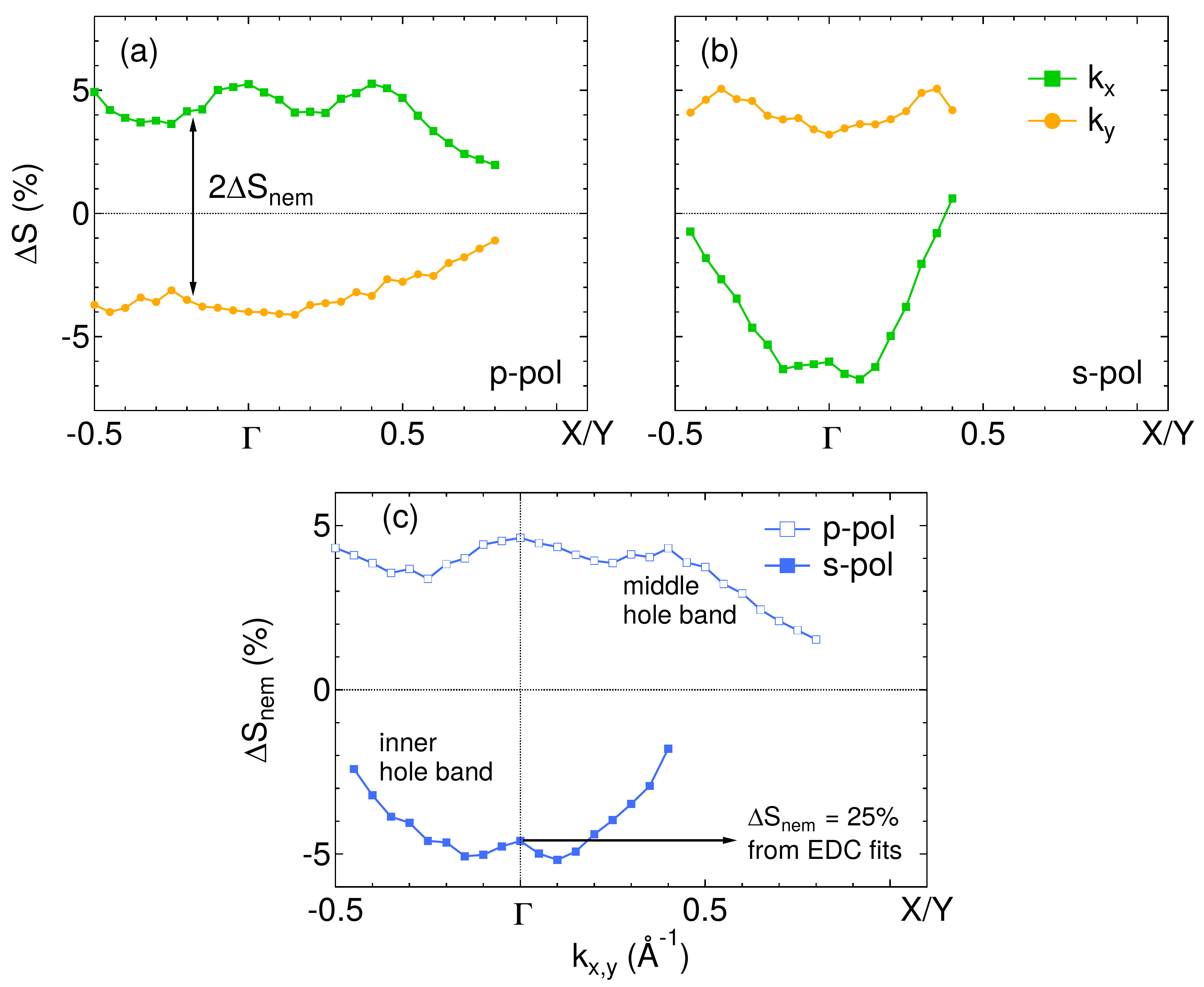}
\caption{
Strain-induced changes in the spectral weight. (a) Difference of the spectral weight  $\Delta S$ for the spectra taken with p-pol light shown in Fig.~\ref{Fig:Ba122}(a,b). (b) same as (a) for s-pol light. (c) Antisymmetric ($B_{2g}$) response of the spectral weight. We indicate the magnitude at $\Gamma$ obtained from EDC fits instead of simple spectral weight integration.
}
\label{Fig:results}
\end{figure}

To quantify the spectral weight change in \BFA~we calculate
\begin{equation}
\Delta S(k_{i}) = 2\frac{ S_{\Delta\epsilon_{yy}>0}(k_{i}) - S_{\Delta\epsilon_{yy}<0}(k_{i}) }  {S_{\Delta\epsilon_{yy}>0}(k_{i}) + S_{\Delta\epsilon_{yy}<0}(k_{i})}~,~i={x,y}
\label{eqn:dS}
\end{equation}
The spectral weight $S_{\Delta\epsilon_{yy}}$ is determined by integration of the intensity $I$ over the whole measured energy range (-0.26,+0.02)\,eV. The results of $\Delta S$  in Fig.~\ref{Fig:results}(a,b) illustrates the almost purely antisymmetric pressure-induced spectral weight change with an opposite sign between $k_x$ and $k_y$ as well as between p-pol and s-pol. We plot this antisymmetric response 
\begin{equation}
 \Delta S_\mathrm{nem}(k) = [\Delta S(k_x) - \Delta S(k_y)] / 2
 \label{eqn:dSnem}
\end{equation}
in Fig \ref{Fig:results}(c). $\Delta S_\mathrm{nem}$ shows that the spectral weight change is largest around $\Gamma$. The spectral weight within the measured energy range is not conserved and is transferred either to larger binding energies or above the Fermi level.

The absolute magnitude of $\Delta S_\mathrm{nem}$ derived from Eqn.~\ref{eqn:dSnem} is underestimated because it includes a large incoherent and strain-independent background in the denominator of Eqn.~\ref{eqn:dS}, which can be seen in the EDCs in Fig.~\ref{Fig:Ba122}. We estimate the magnitude of $\Delta S_\mathrm{nem}$ for the EDCs at $\Gamma$ in s-pol shown in Fig.~\ref{Fig:Ba122}(c3,d3) in two ways. (1) We subtract an energy-independent background with intensity $I(-0.26\,\mathrm{eV})$ and then apply Eqn~\ref{eqn:dS} and \ref{eqn:dSnem}, which gives $\Delta S_\mathrm{nem}=10\%$ as a lower bound. (2) We fit the EDCs with a Lorenzian peak on top of a background, which is composed of an energy-independent constant and a Lorenzian. The function is convoluted with a Gaussian. This leads to $\Delta S_\mathrm{nem}=(25\pm5)$\% as the best estimate. We emphasize that within a linear response regime one expects larger values of $\Delta S_\mathrm{nem}$ for larger applied strains.

\section{Discussion}

In general, the spectral weight anisotropy $\Delta S_\mathrm{nem}$ between the $d_{xz}$ and $d_{yz}$ orbital can originate from a change in admixture from other Fe$3d$ or As$4p$ orbitals. Such a change is expected for any type of nematic order. The relevant Fe $d_{xy}$ and $d_{x^2-y^2}$ orbitals do not contribute to the ARPES spectral weight (Fig.~\ref{Fig:matrix}). Similarly, the overall cross section for As$4p$ is roughly an order of magnitude smaller than for Fe$3d$ \cite{yeh_1985}. Hence, an increase (decrease) of their admixture would decrease (increase) the spectral weight. However, we can clearly exclude Fe$3d$ admixture as the main contribution to $\Delta S_\mathrm{nem}$ because it is symmetry forbidden at $\Gamma$ in both the tetragonal and the orthorhombic state while we observe the largest spectral weight response at $\Gamma$. The As$4p$ orbitals can mix with $d_{xz}$ and $d_{yz}$ also at $\Gamma$. LDA calculations in FeSe show a contribution of Se$4p$ to the inner and middle hole band of approximately 10\% \cite{eschrig_2009_prb} and we expect similar values for \BFA. Only an almost complete transfer of As$4p$ admixture would account for the estimated lower bound of $\Delta S_\mathrm{nem}>10$\%. Therefore, it is unlikely that As$4p$ admixture is the main origin of the spectral weight change at $\Gamma$. 

A nonzero $\Delta S_\mathrm{nem}$ instead suggests a change in quasiparticle coherence. It has been pointed out theoretically that the lifted degeneracy of the $d_{xz}$ and $d_{yz}$ orbitals in the nematic state leads to a different orbital occupation and hence a different quasiparticle coherence within a Hund's metal framework \cite{yu_2018,fanfarillo_2017}. Our experimental results then imply that the $d_{xz}$ orbital becomes more coherent than the $d_{yz}$ orbital. X-ray linear dichroism (XLD) experiments inside the nematic state found a larger occupation of the $d_{xz}$ orbital driving it further away from half filling \cite{kim_2013_prl,rybicki_2020_prb} in agreement with our interpretation. The coherent spectral weight will be redistributed to incoherent Hubbard-like bands located at higher binding energy \cite{stadler_2019}. It is conceivable, that the momentum dependence of $\Delta S_\mathrm{nem}$ with smaller absolute values away from $\Gamma$ is induced by orbital mixture acting on top of Hund's metal physics. 

The resistivity in FeSC is strongly influenced by the degree of quasiparticle coherence determined by their Hund's metal character \cite{haule_2009, medici_2014}. An anisotropic coherence due to nematicity would therefore lead to an anisotropic resistivity as was recently suggested for FeTe. \cite{jiang_2013_prb} The amplitude and sign of the anisotropy will depend on the details of the orbital character distribution around the Fermi surface.

Interatomic interactions, for example due to spin excitations, are believed to be responsible for nematic order. They are expected to induce an anisotropic lifetime \cite{fanfarillo_2016}. It is currently unclear if an anisotropic quasiparticle spectral weight $\Delta S_\mathrm{nem}$ would be another consequence of a bond order.

\section{Conclusion}

In summary, our strain-dependent ARPES studies show that nematicity induces an anisotropic quasiparticle spectral weight in \BFA. The sign of the response implies that the $d_{xz}$ orbital gains spectral weight inside the nematic phase compared to the $d_{yz}$ orbital. We argue that an anisotropic quasiparticle coherence due to Hund's metal physics is compatible with our observation and discuss the role of orbital mixture. Our work offers a momentum and orbital resolved picture of quasiparticle coherence in the nematic state. It highlights the importance of correlation effects inside the nematic state and serves as a test for microscopic theories aiming to explain integrated properties such as resistivity and optical spectroscopy.

\section{Acknowledgments}

\begin{acknowledgments} 
We are grateful for valuable discussions with L. Benfatto, L. Fanfarillo, M. Ikeda and M. Yi. H.P. acknowledges support from the German Science Foundation (DFG) under reference PF 947/1-1. J.C.P. was supported by a Gabilan Stanford Graduate Fellowship and a Stanford Lieberman Fellowship. While writing the paper J.C.P. was supported by a Reines Distinguished Postdoc Fellowship at the National High Magnetic Field Laboratory, Los Alamos National Laboratory. This work was supported by the Department of Energy, Office of Basic Energy Sciences, under Contract No. DE-AC02-76SF00515. Use of the Stanford Synchrotron Radiation Lightsource, SLAC National Accelerator Laboratory, is supported by the U.S. Department of Energy, Office of Science, Office of Basic Energy Sciences under Contract No. DE-AC02-76SF00515. Work at Lawrence Berkeley National Laboratory was funded by the U.S. Department of Energy, Office of Science, Office of Basic Energy Sciences under Contract No. DE-AC02-05-CH11231.
\end{acknowledgments}


\bibliography{manuscript_nem_SW_Ba122}

\end{document}